\documentclass[journal,twoside,web]{ieeecolor}
\usepackage{slashbox}
\usepackage{generic}
\usepackage{cite}
\usepackage{amsmath,amssymb,amsfonts}
\usepackage{algorithmic}
\usepackage{graphicx}
\usepackage{algorithm,algorithmic}
\usepackage{hyperref}
\usepackage{soul}
\usepackage{orcidlink}
\hypersetup{hidelinks=true}
\usepackage{textcomp}
\def\logoname{LOGO-generic-web}
\def\BibTeX{{\rm B\kern-.05em{\sc i\kern-.025em b}\kern-.08em
    T\kern-.1667em\lower.7ex\hbox{E}\kern-.125emX}}
\begin{document}
\title{Time-efficient filtering of polarimetric data by checking physical realizability of experimental Mueller matrices}
\author{Tatiana Novikova \orcidlink{0000-0002-9048-9158}, \IEEEmembership{Member, IEEE}, Alexey Ovchinnikov, Gleb Pogudin, Jessica C. Ramella-Roman
\thanks{This work was supported in part by the European Metrology Programme for Innovation and Research (10.13039/100014132) grant 20IND04 ATMOC (T. Novikova), the SNF Sinergia grant CRSII5 205904 HORAO (T. Novikova) and the NSF grant $\#$1548924 (J. C. Ramella-Roman)} 
\thanks{T. Novikova is with the LPICM, CNRS, Ecole polytechnique, IP Paris, Palaiseau, 91120, France. She is also associated  with the Department of Biomedical Engineering, Florida International University, Miami, FL 33174, USA (e-mail: tatiana.novikova@polytechnique.edu).}
\thanks{A. Ovchinnikov is with the Department of Mathematics, Queens College, City University of New York, Queens, NY 11367-1597 USA (e-mail: alexey.ovchinnikov@qc.cuny.edu).}
\thanks{G. Pogudin is with the Laboratoire d’informatique de l’École Polytechnique, IP Paris, Palaiseau, 91120, France (e-mail: gleb.pogudin@polytechnique.edu)}
\thanks{J. C. Ramella-Roman is with the Department of Biomedical Engineering, Florida International University, Miami, Florida 33174 USA. She is also associated with the Herbert Wertheim College of Medicine, Florida International University, Miami, Florida 33199 USA, (e-mail: jramella@fiu.edu)}
\thanks{The names of the authors are listed alphabetically and all authors share equal credit for the paper.}
}

\maketitle

\begin{abstract}
Imaging Mueller polarimetry has already proved its potential for metrology, remote sensing and biomedicine. The real-time applications 
of this modality 
require both video rate image acquisition and fast data post-processing algorithms. 
First, one must check the physical realizability of the experimental Mueller matrices in order to filter out non-physical data, i.e. to test the positive semi-definiteness of the $4\times4$  Hermitian coherency matrix calculated from the elements of the corresponding Mueller matrix pixel-wise.  
For this purpose, we compared the execution time 
for the calculations of i) eigenvalues, ii) Cholesky decomposition, iii) Sylvester's criterion, and iv) coefficients of the characteristic polynomial of the Hermitian coherency matrix using two different approaches, all calculated for the experimental 
Mueller matrix images (600 pixels$\times$700 pixels) of mouse uterine cervix.  
The calculations were performed using  C++ and Julia programming languages. Our results showed the superiority of the algorithm iv), in particular, the version based on the simplification via Pauli matrices,  
in terms of 
execution time  
for our dataset, over other algorithms. The sequential implementation of the latter algorithm on a single core already satisfies the requirements of real-time polarimetric imaging in various domains.
This can be further amplified by the proposed parallelization (for example, we achieve a 5-fold speed up on 6 cores). 
\end{abstract}


\section{Introduction}
\label{sec:introduction}
\IEEEPARstart{M}{ueller} polarimetry (MP) is an optical technique that is used for the characterization of polarimetric properties of a sample for various applications such as metrology \cite{
NovBul2021, Liu2015, J3M}, remote sensing \cite{Kat1999, 
HM2023},  medical imaging \cite{ Lee2021SciRep, BookBrainChapter, Vit2015, Pierangelo2011, Travis}, tissue engineering \cite{photonics10101129,BR}, digital histology \cite{LeeDBSCAN, MaHuiDH, Ush2021} material characterization \cite{MaterialC,  Arteaga:19}, food quality control \cite{app12105258}, etc. By measuring the changes in polarization states of incident light upon interaction with a sample a MP system provides a set of 16 parameters that form a $4\times4$ real-valued transfer matrix (so-called Mueller matrix (MM)) \cite{Gold2010, Azzam1978, GRO} of this sample. 
Each element of MM corresponds to a specific transformation of the input polarization state. Choosing an appropriate MM decomposition algorithm \cite{Lu1996, Cloude, Li2020, photonics6010034, GilPhotonics, Ghosh2009}  allows researchers to extract and analyze quantitatively the depolarization, retardance and diattenuation properties \cite{Gold2010} of a sample which carry relevant metrological (or diagnostic) information. 

The design and calibration of MM polarimeters are usually optimized in terms of minimizing the measurement errors \cite{Tyo, DeMartino, Bruce, Compain}. However, the presence of the residual systematic and random experimental errors may lead to nonphysical measurement results at certain measurement wavelengths or/and at certain image pixels, i. e. produce a particular $4\times4$ real-valued matrix that does not obey the condition of physical realizability for MM (see Sec.\ref{sec:2}). It is worth to mention that a necessary and sufficient condition for the physical realizability of
Mueller matrix 
requires the corresponding $4\times4$ Hermitian coherency matrix (HCM) \cite{Gold2010} to be a positive semi-definite matrix \cite{Cloude, CloudeSPIE}. Thus, checking this condition for all acquired MMs represents a must step of the polarimetric data post-processing for filtering out nonphysical data.

In particular, the imaging MM polarimeters operating in either real \cite{photonics10101129, Lee2021SciRep,BookBrainChapter} or reciprocal \cite{J3M} space produce 
a significant amount of polarimetric data
at each single measurement, because the total number of image pixels is quite large (the resolution of CCD cameras is typically several hundreds of thousands pixels or more). Consequently, for the real-time imaging applications with a tight time budget it is important to find the most time-efficient algorithm for checking pixel-wise whether a $4\times4$ HCM is a positive semi-definite one.

The paper is organized as follows: in the next section we describe the basics of Stokes-Mueller formalism and the 
MM images of mouse uterine cervix taken 
with the custom-built MM polarimetric system
and used in our studies. Four different approaches for testing the physical realizability of the measured MMs, namely, the calculations of i) eigenvalues, ii) Cholesky decomposition, iii) Sylvester's criterion
, and iv) coefficients of the
characteristic polynomial of the corresponding HCM and the comparison of these algorithms in terms of the execution time
and accuracy  are presented and discussed in the third section of the paper. The last section summarizes our results and discuss the perspectives of potential applications.

\section{Materials and Methods}
\subsection{Stokes-Mueller formalism}
\label{sec:2}
In this section we briefly recall the theoretical framework for the description of fully or partially polarized light. 

Apart from light intensity and color the polarization of light reflects its vectorial nature. Let us consider the propagation of a plane polarized monochromatic electromagnetic (EM) wave through an infinite isotropic medium.  By selecting a Cartesian coordinate system with $Oz$ axis parallel to the direction of wave propagation ($\vec{k} = k\hat{\vec{z}}$, where $k$ is a wave number) the oscillation of the electric field (EF) vector $\vec{E}$ at the point $(0, 0, z)$ is confined to the plane 
$Ox$-$Oy$ 
orthogonal to $Oz$ axis and is described by a superposition of two independent harmonic oscillators  \cite{Gold2010}
\begin{equation}
\label{Eq:1}
\begin{split}
E_x (z,t)=E^0_{x}\cos (\omega t-kz+\delta_x ) \\
E_y (z,t)=E^0_{y} \cos(\omega t-kz+ \delta_y)
\end{split}
\end{equation}
where $E^0_{x}$ and $E^0_{y}$ are the constant amplitudes, $\omega$ is the angular frequency, $k$ is the wavenumber, $\delta_x$ and $\delta_y$ are the arbitrary constant phases, and the subscripts $x$ and $y$ refer to the field components in the $x$- and $y$-directions, respectively. Equations \eqref{Eq:1} can be re-written as
\begin{equation}
\label{Eq:2}
\left(\frac{E_x}{E^0_{x}}\right)^2 +\left(\frac{E_y}{E^0_{y}}\right)^2 -2 \frac{E_x E_y}{E^0_{x} E^0_{y}} \cos \delta = \sin^2 \delta
\end{equation}
where $\delta = \delta_y-\delta_x$ is the phase shift between the orthogonal transverse components of EF vector of a plane EM wave. Equation \eqref{Eq:2} describes a \textit{polarization ellipse} that represents the trajectory of a tip of oscillating EF vector of polarized light. However, the direct observation of the polarization ellipse is impossible, because optical EF vector traces out the locus of points described by \eqref{Eq:2} in a time interval of order $10^{-15}$ s.

For partially or completely depolarized light the motion of EF vector is disordered in $Ox$-$Oy$ plane and can be described by its probability distribution only. We cannot observe the amplitude of optical EF, but we can observe and measure the intensity of light. For linear optical systems the measured intensity is obtained by
averaging the square of EF amplitude and represents 
the second moments (quadratic quantities) of the EF distribution. 

We denote the intensities related to the components of the EF vector parallel to $Ox$ and $Oy$ axes as $I_x=\langle E_x^2\rangle$, $I_y=\langle E_y^2\rangle$, respectively. The angle brackets $\langle\cdot\rangle$ indicate temporal, spatial and spectral averaging,
which depends on both sample and measurement conditions. 
We define four observable Stokes parameters as
\begin{equation}
\label{Eq:3}
\begin{split}
S_0 &=0.5\langle E_x^2+E_y^2 \rangle \\
S_1 &= 0.5\langle E_x^2 - E_y^2 \rangle \\
S_2 &=\langle E_x^2 E_y^2 \cos \delta \rangle \\
S_3 &=\langle E_x^2 E_y^2 \sin \delta \rangle
\end{split}
\end{equation}
The parameter $S_0$  is the total intensity of light, $S_1$ represents the difference in the light intensities 
measured after a linear polarizer with optical axis aligned parallel with either $Ox$ or $Oy$ axes. The parameter $S_2$ is equivalent to the   parameter $S_1$, but the intensities are measured after a linear polarizer with optical axis aligned at either $+45^\circ$ or $-45^\circ$, respectively, within the plane orthogonal to the direction of light beam propagation. 
The parameter $S_3$ represents the difference between 
the intensities transmitted by either left or right circular polarizer. 

We define the Stokes vector as $\vec{S}=(S_0, S_1, S_2, S_3)^T$
The degree of polarization $\rho$ of any Stokes vector $\vec{S}$ is defined as

\begin{equation}
\label{Eq:4}
\rho=\frac{\sqrt{S_1^2+S_2^2+S_3^2}}{S_0}, \quad (0 \leq \rho \leq 1)
\end{equation}

\noindent where parameter $\rho$ varies between 0, for totally depolarized light, and 1, for totally polarized light.
Upon interaction with a sample the Stokes vector of incident light $\vec{S}^{in}$ undergoes a linear transformation, described by a $4 \times 4$ real-valued matrix $\mathbf{M}$, which is called a Mueller matrix (MM) \cite{Gold2010} of a sample. 
\begin{equation}
\label{Eq:5}
\begin{bmatrix}
S_0^{out} \\
S_1^{out} \\
S_2^{out} \\
S_3^{out} \\
\end{bmatrix}
=
\begin{bmatrix}
m_{00} &  m_{01}& m_{02} &  m_{03} \\
m_{10} &  m_{11}& m_{12} &  m_{13}  \\
m_{20} &  m_{21}& m_{22} &  m_{23}  \\
m_{30} &  m_{31}& m_{32} &  m_{33} \\
\end{bmatrix}
\begin{bmatrix}
S_0^{in} \\
S_1^{in} \\
S_2^{in} \\
S_3^{in} \\
\end{bmatrix}
\end{equation}

It is obvious that physically realizable MM should transform any Stokes vector of incident light beam into a Stokes vector of output light beam with the degree of polarization less than or equal to 1. 
An Hermitian $4\times4$ 
coherency matrix $\mathbf{H}$ associated with matrix $\mathbf{M}$ is defined as in~\cite{GRO}:
\begin{equation}\label{eq:Hpauli}
\mathbf{H} = \frac{1}{4}\sum\limits_{i, j = 0}^3 m_{i j} (\sigma_i \otimes \sigma_j^T),
\end{equation}
where $\otimes$ is the Kronecker product of matrices and $\sigma_0, \ldots, \sigma_3$ is an extended set of Pauli matrices \cite{Pauli}:
\[
\sigma_0 = \mathbf{I}_2, \sigma_1 = \begin{pmatrix}
    1 & 0 \\
    0 & -1
\end{pmatrix}, \sigma_2 = \begin{pmatrix}
    0 & 1 \\
    1 & 0
\end{pmatrix}, \sigma_3 = \begin{pmatrix}
    0 & -i\\
    i & 0
\end{pmatrix}.
\]

It was demonstrated by Cloude~\cite{Cloude, CloudeSPIE} that a $4\times 4$ real-valued matrix $\mathbf{M}$ represents a physically realizable MM if and only if the associated HCM $\mathbf{H}$ calculated in \eqref{eq:Hpauli} is a positive semi-definite matrix. 
Due to the presence of measurement systematic errors and noise, this condition may not always be satisfied for the experimental polarimetric data. Thus, these data must undergo a physical realizability filtering before further data post-processing.  

\subsection{Mueller matrix images of mouse uterine cervix}
Imaging Mueller polarimetry has proven its potential for various biomedical applications by helping the clinicians to make better decisions about patient diagnosis and treatment \cite{Pierangelo2011,
Vizet2017, Ji}. The translation 
of this modality 
into clinical practice requires 
video rate image acquisition and the development and implementation of fast polarimetric data post-processing algorithms. As was mentioned above the pixel-wise test of physical realizability of MM images is a first step of polarimetric data post-processing.

\begin{figure}[!h]
\centerline{\includegraphics[width=\columnwidth]{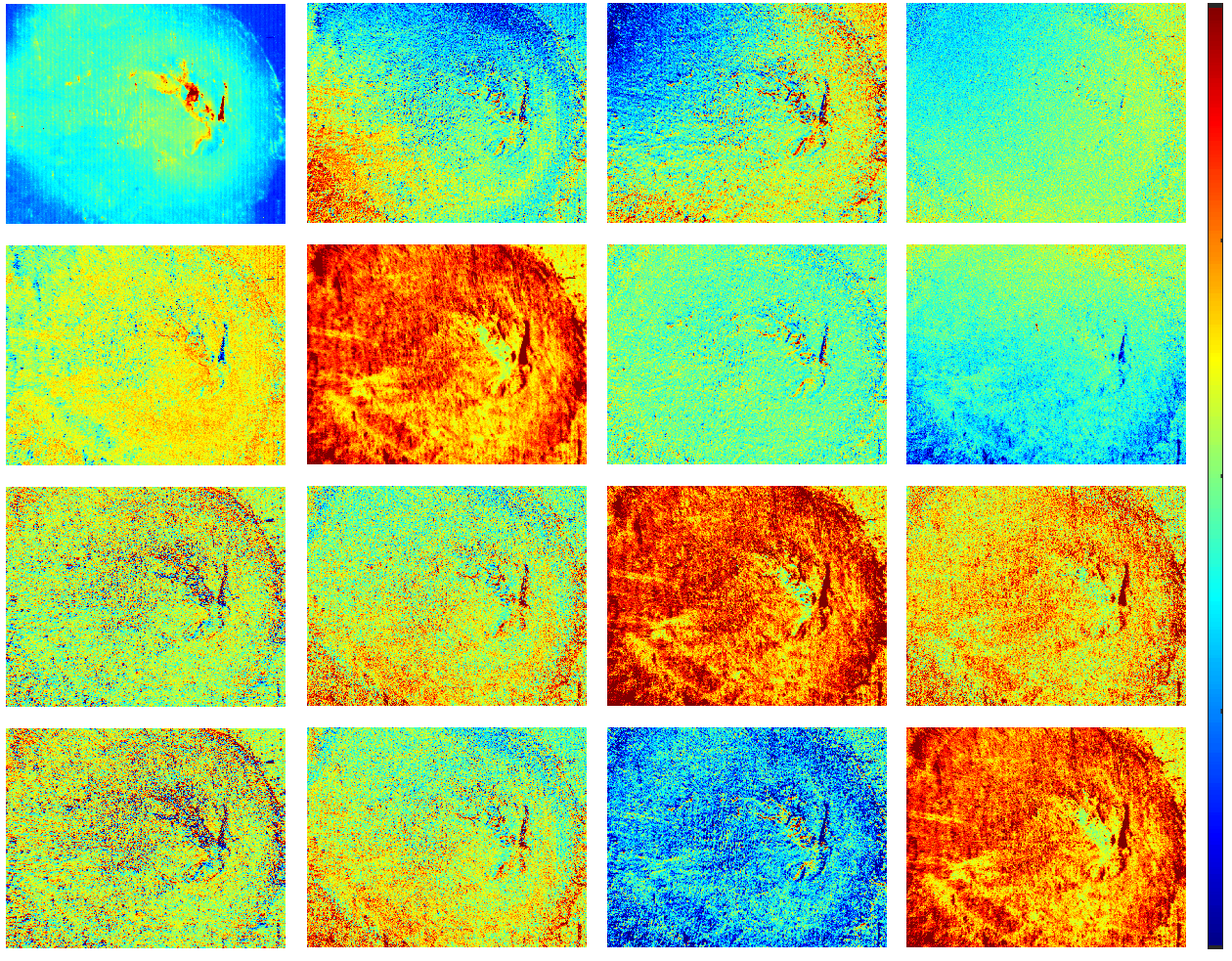}}
\caption{ 
Mueller matrix images  
of 
mouse uterine cervix taken
with a Mueller polarimeter.  
All elements of Mueller matrix (except of $m_{00}$) are normalized by $m_{00}$ values pixel-wise. The color bar values vary within i) $[0;0.5]$  for $m_{00}$, ii) $[0;0.1]$  for all off-diagonal elements, 
and 
iii) $[0;0.2]$  for the elements $m_{11}$, $m_{22}$, $m_{33}$.
The image size is 
2 mm
$\times$ 
2.5 mm.}
\label{fig1}
\end{figure}

In our studies we used the experimental 
Mueller matrix images (600 pixels $\times$ 700 pixels) of thin section (nominal thickness 50 $\mu m$) of
mouse uterine cervix (Fig.\ref{fig1}). The research protocols were reviewed and approved by the Florida International University (registration number: IACUC-20-014).

These images were acquired
with the custom-built imaging Mueller polarimetric system described in \cite{TN_JRR}. 
The modulation of incident light polarization and analysis of the detected signal
was performed with the rotating quarter-wave plates and fixed polarizers inserted into both illumination and detection arm of the instrument.  
A 550-nm LED
(Thorlabs, Newton, New Jersey) was used as a light source, a sCMOS camera (pco.edge by PCO, Kelheim, Germany) was used as a detector for the image registration.
The system calibration was done
with the eigenvalue calibration method described by Compain 
\cite{Compain}.
The experimental errors on the elements of measured Mueller matrix of air (it should be a unity matrix) were below 1\%. The complete description of the experimental setup can be found in \cite{TN_JRR}. 

During data post-processing step the check on physical realizability of each matrix from the dataset containing 420000 ($=600\times700$) matrices has to be performed first. Given the massive amount of polarimetric data to be tested one needs to find the most time-efficient algorithm for checking the physical realizability of large dataset of MMs (i. e.
positive semi-definiteness of the corresponding HCMs).
\section{Tests on positive semi-definiteness of 
HCM}
\subsection{Algorithms}
\label{Sec:3}
\noindent \textbf{Definition:} Hermitian $n\times n$ matrix $\mathbf{H}$ is positive semi-definite (or definite) if  $(\mathbf{H}\vec{x}, \vec{x})\geq0$ (or $>0$) for all $\vec{x} \in \mathbb{C}^n \backslash \{\vec{0}\}$. 

We have tested four different approaches for checking the positive semi-definiteness (or positive definiteness) of HCM by calculating:
\begin{itemize}
    \item eigenvalues of HCM,
    \item Cholesky decomposition of HCM,
    \item Sylvester's criterion for HCM, 
    \item  coefficients of HCM characteristic polynomial.
\end{itemize}
It is worth considering both the pros and cons of each algorithm for filtering out nonphysical polarimetric data.
\subsubsection{HCM eigenvalues}
Nowadays, the most common test for the physical realizability of a MM relies on the calculation of eigenvalues of the corresponding HCM \cite{Cloude, Gil, CloudeSPIE}. All eigenvalues are nonnegative for a positive semi-definite matrix. Moreover, if MM of a sample is partially or completely depolarizing, the corresponding HCM must be a positive definite matrix. The latter comment is particularly relevant for almost all types of biological tissue because of strong light scattering within biotissue leading to the significant depolarization of detected scattered light \cite{Tuchin}. 

This method is widely spread, because one may not only filter out nonphysical data but also use a vector measure of depolarization based on the three smallest nonnegative eigenvalues as  of the physically realizable MM. It was shown that such a metric may increase the polarimetric image contrast for certain classes of samples \cite{rs14174148, GilPhotonics, Shepard}. However, for a wide class of samples, it is not necessary to know the exact eigenvalues of HCM, 
when one needs
to evaluate the sign of eigenvalues only.

\subsubsection{Cholesky decomposition of HCM}
A decomposition of a Hermitian positive definite matrix $\mathbf{H}$ into the product of a lower triangular matrix $\mathbf{L}$ and its conjugate transpose matrix $\mathbf{L}^*$ is called Cholesky decomposition: $\mathbf{H}=\mathbf{L}\mathbf{L}^*$, and Cholesky decomposition is unique for such matrix \cite{Haddad2009}. To test whether a Hermitian matrix is positively definite using Cholesky decomposition, one needs to check whether the signs of all diagonal elements of the lower triangular matrix are positive. The fact that Cholesky decomposition may address the HCM positive definiteness only is not very restrictive for medical applications of MP, because, in general, biological tissue are highly depolarizing media and the corresponding HCMs are positive definite.   
\subsubsection{Sylvester's criterion
} 
\begin{figure}[!h]
\centerline{\includegraphics[width=0.37\columnwidth]{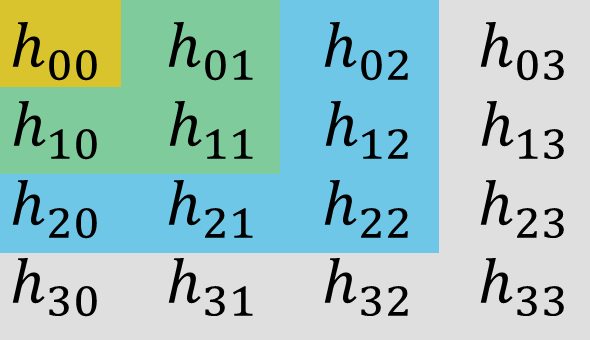}}
\caption{Illustration of the Sylvester's criterion: calculations of leading principal minors of HCM, which are the determinants of the top left corner sub-matrices rendered in orange ($1\times1$), orange+green ($2\times2$), orange+green+blue ($3\times3$), and orange+green+blue+gray ($4\times4$) colors}
\label{fig2}
\end{figure}

Sylvester's criterion of positive definiteness of an arbitrary $n\times n$ Hermitian matrix states that this matrix is positive definite if and only if all its leading principal minors are positive \cite{Gil2000, Gilbert, CloudeSPIE }. In case of a $4\times 4$ HCM it means that all following matrices have a positive determinant:
\begin{itemize}
   \item 
    upper left $1\times1$ sub-matrix of HCM,
    \item 
   upper left $2\times2$ sub-matrix of HCM,
    \item 
upper left $3\times3$ sub-matrix of HCM, 
    \item 
    HCM itself.
\end{itemize}
The choice of corresponding elements of HCM for the calculations of leading principal minors is illustrated in Fig.\ref{fig2}.
In the actual implementation,  we applied (in advance) {\sc Maple}'s command {\tt simplify} to each of the determinant expressions to decrease the computing time.
\subsubsection{Coefficients of HCM characteristic polynomial}
The characteristic polynomial (CP) of arbitrary $n\times n$ matrix $\mathbf{H}$ is defined as function $f(\lambda)= \det(\mathbf{H} + \lambda \mathbf{I}_n)$, where $\mathbf{I}_n$ represents the $n\times n$ identity matrix \cite{BROOKS} (here we used ``$+$'' instead of the standard ``$-$'' in the definition of CP so that it will be easier to state the criterion). 
An arbitrary $n\times n$ matrix $\mathbf{H}$ with real eigenvalues is positive semi-definite if and only if all coefficients of its characteristic polynomial are nonnegative. 
This is true by applying Descartes' rule of signs~\cite{Wang2004} to $f(\lambda)$ and using the equivalence between positive semi-definiteness and non-negativity of the eigenvalues.  

It turns out that the actual presentation of the formulas for the coefficients of $f(\lambda)$ can affect the computing time. In what follows, we use two versions for the presentation of the coefficients:
\begin{itemize}
\item formulas separately simplified in {\sc Maple} using command {\tt simplify}
\item formulas obtained using Pauli matrices \cite{Pauli} 
as described below.
\end{itemize}

The coefficients of the CP of $\mathbf{H}$ are the elementary symmetric polynomials of its eigenvalues $\lambda_1, \ldots, \lambda_4$. 
We could find the values of these polynomials using the Newton identities~\cite[p.~23]{Macdonald} if we knew the values of the power sums $\lambda_1^j + \ldots + \lambda_4^j$ for $j=1, \ldots, 4$.
These power sums, in turn, are exactly the traces $\operatorname{tr}\mathbf{H}, \ldots, \operatorname{tr}(\mathbf{H}^4)$.
For computing these traces, we will use the following observations:
\begin{itemize}
    \item Any power of $\mathbf{H}$ is again a linear combination of the form~\eqref{eq:Hpauli};
    \item $\operatorname{tr}(\sigma_i \otimes \sigma_j^T) = 4$ if and only if $i = j = 0$, and it is equal to $0$ otherwise.
\end{itemize}
The latter property immediately gives us a simple formula $\operatorname{tr}\mathbf{H} = m_{00}$.
For the second degree, we observe that the only way to obtain $\mathbf{I}_2$ as a product of two Pauli matrices is to take a square. This implies that
\[
\operatorname{tr}(\mathbf{H}^2) = \frac{1}{16}\sum\limits_{i, j = 0}^3 m_{ij}^2.
\]
Using similar but much more tedious combinatorial considerations, we have obtained formulas for $\operatorname{tr}(\mathbf{H}^3)$ and $\operatorname{tr}(\mathbf{H}^4)$, presented in the Appendix. 

\subsection{Comparison of the results: execution time and accuracy}
All  algorithms described in Sec. \ref{Sec:3} were tested on the experimental Mueller matrix data described in Sec. \ref{sec:2}. The results of the comparison in terms of the execution time depending the number of used CPU cores are shown in Tab.~\ref{Tab:1}. The code was written in C++, and the experiments were performed on a Mac computer with CPU 3.1 GHz 6-Core Intel Core i5 and 16 GB. 
For the eigenvalue computation and Choletsky decomposition we used the numerical linear algebra library
    Eigen~\cite{eigenweb} (version 3.4.0). 
    The C++ compiler was Clang 13.1.6, and the Mac OS version was 12.3. The runtimes were averaged over 20 runs; the standard deviation was $1.6\%$ or lower. The implementation is available here: \url{https://github.com/pogudingleb/mueller_matrices}.
\begin{table*}[h!]
\caption{Execution time in milliseconds for C++ implementation of the  algorithms from \\Section~\ref{Sec:3} checking whether 420,000 Mueller matrices are physically realizable}
\label{Tab:1}
\normalsize
\centering
\setlength{\tabcolsep}{3pt}
\begin{tabular}{|c|r|r|r|r|r|r|}
\hline
\backslashbox{Algorithm}{Used CPU cores}& 
1& 
2 &3&4&5&6\\
\hline
Eigen's self-adjoint eigensolver  & 561.3 & 289.1 & 194.3 & 145.5 & 119.2 & 99.5\\
Eigen's Cholesky decomposition  & 49.6 & 25.5 & 17.1 & 13.0 & 10.6 & 8.8 \\
Sylvester's criterion & 34.6 &   17.7 &  11.9 &  9.0 &  7.4 &  6.3 \\
Coefficients of CP (via {\sc Maple} {\tt simplify})
& 26.8 & 13.5& 9.1 & 6.9 & 5.6 & 4.7
 \\
 Coefficients of CP (simplification via Pauli matrices) 
& 13.1 & 6.6 & 4.5 & 3.5 & 2.9 & 2.6
 \\
\hline
\end{tabular}
\label{tab1}
\end{table*}

The number of filtered out nonphysical matrices is 868 and
constitutes approximately $0.2\%$ of the total number of matrices for all tested algorithms, thus, proving the good quality of our experimental polarimetric data. In addition, all nonphysical data were found at the same image pixels 
by all algorithms. 
So, all the algorithms were numerically stable enough for the considered data.

Our computational experiments confirm that the method of calculations of HCM eigenvalues for checking positive  definiteness (semi-definiteness) of HCM is orders of magnitude slower and more memory demanding compared to other three methods.

We achieve additional speed-up by parallelization. Indeed, the matrix for each pixel can be processed independently of the other pixels. On our 6-core environment we achieved a run-time speed-up of approximately 5.5 times over the sequential computation. This is more than enough  for the processing of the matrices to be used in real-time.

For comparison, we also ran this in Julia version 1.10.0 in CentOS Linux 7, on Intel(R) Xeon(R) CPU E5-2680 v2 @ 2.80GHz, with 20 cores, and found out that the relative placement of the algorithms in terms of runtime is the same as that of our C++ code. Our Julia implementations of the algorithms via Sylvester's criterion and via coefficients of the CP are also sufficiently fast to be used in real time. The detailed results are reported here: \url{https://github.com/pogudingleb/mueller_matrices/blob/main/src_julia/julia_code.out}.

\section{Conclusion}
Our results on
checking pixel-wise physical realizability of the experimental Mueller matrix images of mouse uterine cervix 
demonstrated the superiority
of two algorithms (Sylvester's criterion and calculations of the coefficients of the characteristic polynomial of HCM)
for our dataset,
in terms of
execution time 
compared to other considered algorithms, whereas keeping the same accuracy. 
It is quite logical that the filtering method based on the calculations of HCM eigenvalues may not be optimal 
because we generate the redundant data, i. e. the absolute values of the eigenvalues of HCM, that are not used for MM filtering. 

Another explanation for the observed performance gap is related to the fact that general-purpose linear algebra software is often not optimized for the matrices of small dimensions (e. g. $4\times4$), which
we consider in this paper.

Our findings are important for many applications of MP that generate significant amount of polarimetric data and require both real-time data acquisition and fast data
post-processing algorithms, in particular, for imaging MP used for i) metrology in semi-conductor industry, and ii) \emph{in vivo} medical diagnosis and surgery guidance in clinical settings. 

\section*{Appendix}

In this appendix, we will provide explicit formulas for $\operatorname{tr}(\mathbf{H}^3)$ and $\operatorname{tr}(\mathbf{H}^4)$, which we obtained through combinatorial considerations with Pauli matrices.
Let $\widetilde{\mathbf{M}}$  denote the lower-right $3\times 3$-submatrix of $\mathbf{M}$.

We start with $\mathbf{H}^3$.
We have
\begin{multline*}
\operatorname{tr}(\mathbf{H}^3) = \frac{3}{4}m_{00}\operatorname{tr}(\mathbf{H}^2) - \frac{m_{00}^3}{8} \\- \frac{3}{8}\left(\operatorname{det}\widetilde{\mathbf{M}} + \sum\limits_{i = 1}^3 m_{i, 0}\sum\limits_{j = 1}^3 m_{0j}m_{ij}\right).
\end{multline*}
For $\operatorname{tr}(\mathbf{H}^4)$, we will introduce some intermediate variables.
Let $S_i := \sum\limits_{j = 1}^3 m_{ij}^2$ for $i = 1, 2, 3$, and define
\begin{align*}
    A &= \sum\limits_{i = 1}^3 S_i^2,\\
    B &= m_{10}^2(S_2 + S_3) + m_{20}^2 (S_1 + S_3) + m_{30}^2 (S_1 + S_2).
\end{align*}
Then, for $i < j < 4$, we set
\[
P_{ij} = m_{i0}m_{j0} - (-1)^{\delta_{i0}}\sum\limits_{k = 1}^3 m_{ik}m_{jk}
\]
and define
\[
C = -P_{01}^2 - P_{02}^2 - P_{03}^3 + P_{12}^2 + P_{13}^2 + P_{23}^2.
\]
We also define
\[
D = \sum\limits_{i = 1}^3 m_{0i}^2, \quad F = \sum\limits_{i = 1}^3 m_{i0}^2.
\]
Then the final formula will be:
\begin{multline*}
\operatorname{tr}(\mathbf{H}^4) = \frac{-\det(\mathbf{M})}{8} + \frac{3}{4}\left(\operatorname{tr}(\mathbf{H}^2)\right)^2 - \frac{A}{32} - \frac{B}{16}  \\ - \frac{D}{16}\left(\frac{F}{2} + S_1 + S_2 + S_3\right) \\+ \frac{m_{00}}{4}\left(2\operatorname{det}(\widetilde{\mathbf{M}}) + \sum\limits_{i = 1}^3 m_{i, 0}\sum\limits_{j = 1}^3 m_{0j}m_{ij}\right)  \\ - \frac{C}{16} - \frac{m_{00}^2F}{16} - \frac{1}{32}\left(m_{00}^4 + m_{10}^4 + m_{20}^4 + m_{30}^4\right).
\end{multline*}
The automatic verification of the formulas performed using {\sc Maple} can be found in the repository\footnote{\url{https://github.com/pogudingleb/mueller_matrices/blob/main/formula.mpl}}.

\section*{Acknowledgments}
We thank Georgy Scholten for fruitful discussions and valuable advice. We also thank the Courant Institute of Mathematical Sciences for its computational resources. None of the authors declare
any conflict of interest. 

\section*{Data availability}
The experimental Mueller matrix data and 
codes for all algorithms used in the current study are available in the  repository: 
\url{https://github.com/pogudingleb/mueller_matrices}.

\section*{References}
\bibliographystyle{unsrt}
\bibliography{Biblio}
\end{document}